\documentclass[aps,10pt,reprint,groupedaddress,superscriptaddress]{revtex4-2}

\usepackage[usenames,dvipsnames]{xcolor}
\usepackage{amssymb}
\usepackage{graphicx}
\usepackage{amsmath}
\usepackage{hyperref}
\usepackage{braket}
\usepackage[normalem]{ulem}
\definecolor{pink2}{rgb}{0.858, 0.188, 0.478}
\usepackage{placeins}
\DeclareMathOperator{\Tr}{Tr}

\begin{document}
\title{Observing emergent  hydrodynamics in a long-range quantum magnet}


\newcommand{\TUM}{\affiliation{Department of Physics and Institute for Advanced Study, Technical University of Munich, 85748 Garching, Germany}}
\newcommand{\MCQST}{\affiliation{Munich Center for Quantum Science and Technology (MCQST), Schellingstr. 4, 80799 M{\"u}nchen, Germany}}
\newcommand{\IQOQI}{\affiliation{Institute for Quantum Optics and Quantum Information, Austrian Academy of Sciences, Technikerstra{\ss}e 21a, 6020 Innsbruck, Austria}}
\newcommand{\UIBK}{\affiliation{Institut f\"ur Experimentalphysik, Universit\"at Innsbruck, Technikerstra{\ss}e 25, 6020 Innsbruck, Austria}}
\newcommand{\AQT}{\affiliation{AQT, Technikerstra{\ss}e 17, 6020 Innsbruck, Austria}}

\author{M.~K.~Joshi} \IQOQI
\author{F.~Kranzl} \IQOQI \UIBK
\author{A. Schuckert} \TUM \MCQST
\author{I. Lovas} \TUM \MCQST
\author{C. Maier} \IQOQI \AQT
\author{R. Blatt} \IQOQI \UIBK
\author{M. Knap} \TUM \MCQST
\author{C.~F.~Roos} \IQOQI \UIBK

\begin{abstract}

Identifying universal properties of non-equilibrium quantum states is a major challenge in modern physics. A fascinating prediction is that classical hydrodynamics emerges universally in the evolution of any interacting quantum system. Here, we experimentally probe the quantum dynamics of 51 individually controlled ions, realizing a long-range interacting spin chain. By measuring space-time resolved correlation functions in an infinite temperature state, we observe a whole family of hydrodynamic universality classes, ranging from normal diffusion to anomalous superdiffusion, that are described by L\'evy flights. We extract the transport coefficients of the hydrodynamic theory, reflecting the microscopic properties of the system. Our observations demonstrate the potential for engineered quantum systems to provide key insights into universal properties of non-equilibrium states of quantum matter. 
\end{abstract}

\maketitle


Universality in equilibrium asserts that microscopic details are irrelevant for the emergent quantum phases of matter and their transitions. Rather, symmetries and topology determine the essential macroscopic properties. By contrast, all scales, from low to high energies, are relevant for quantum systems which are driven far from their thermal equilibrium. Recent experimental progress in engineering coherent and interacting quantum systems made it possible to create and explore exotic non-equilibrium states, which can exhibit unconventional relaxation dynamics~\cite{gring_relaxation_2012, Bohnet:2016, Tang_2018, bernien_probing_2017, guardado-sanchez_2020, scherg_2020}, dynamical phases~\cite{schreiber_observation_2015, Smith_2016, kaufman_quantum_2016, Brydges_2019, prufer_observation_2018, erne_universal_2018}, and transitions between them~\cite{PhysRevLett.119.080501, monroeDQPT}. 

Despite this wealth of observed quantum phenomena, a common anticipation is that classical hydrodynamics of a few conserved quantities emerges universally for any complex quantum system, as strong interactions entangle and effectively mix local degrees of freedom~\cite{lux_hydrodynamic_2014, Bohrdt_2017}. However, verifying this assumption, and furthermore determining the non-universal transport coefficients of the emergent hydrodynamic theory for specific systems is challenging. 
Recently, enormous efforts have been devoted to detect hydrodynamic transport in quantum gases~\cite{cao_universal_2011, sommer_universal_2011, schneider_fermionic_2012, brown_bad_2019} and condensed matter systems~\cite{Bandurin_2016, Crossno_2016, Moll_2016, zu_2021}. While transport is generally expected to be diffusive, a variety of largely unexplored classes of hydrodynamics have been  predicted theoretically including anomalous subdiffusive~\cite{Gromov_2020, feldmeier_2020a} and superdiffusive transport~\cite{Ljubotina_2017,bulchandani_2021, PhysRevB.101.020416}. 

In this work, we experimentally probe the dynamics of a long-range quantum magnet, realized in a quantum simulator of 51 individually controlled ions. We develop a protocol to measure space- and time-resolved spin correlations in an engineered infinite temperature state, enabling us to experimentally establish that hydrodynamics emerges in the non-equilibrium quantum state. By tuning the long-range character of the interactions, we observe a whole family of hydrodynamical universality classes, ranging from normal diffusion to anomalous superdiffusion. 

The (pseudo-)spins of our quantum system are realized with two electronic states of the $^{40}$Ca$^+$ ion: $\ket{S_{1/2},m=+1/2}$ as $\ket{\downarrow}$ and  $\ket{D_{5/2},m=+5/2}$ as $\ket{\uparrow}$. The quantum state of individual ions is controlled by a tightly focused, steerable laser beam capable of addressing any ion in the string, in conjunction with a laser beam that collectively interacts with the ions.
A two-tone laser field realizes approximately power-law decaying Ising interactions between the (pseudo-)spins by off-resonantly coupling motional and electronic degrees of freedom of the ion chain.
Application of a strong transverse field energetically penalizes spin non-conserving contributions~\cite{jurcevic_quasiparticle_2014}. The effective dynamics is then described by the long-range XY model
\begin{equation}
    \hat H= \sum_{i<j} \frac{J}{|i-j|^\alpha} (\hat \sigma^+_i\hat\sigma^-_j + \hat \sigma^-_i\hat\sigma^+_j),
    \label{eq:xy}
\end{equation}
where $J$ and $\alpha$ determine the strength and the range of the spin-spin interaction, respectively \cite{Monroe:2021}. They can be tuned by varying the amplitude and frequency of the two-tone laser field. Further experimental details are given in the supplementary materials \cite{SI}, Sec.~I-IV. 

\begin{figure*}[ht!]
\centering
\includegraphics{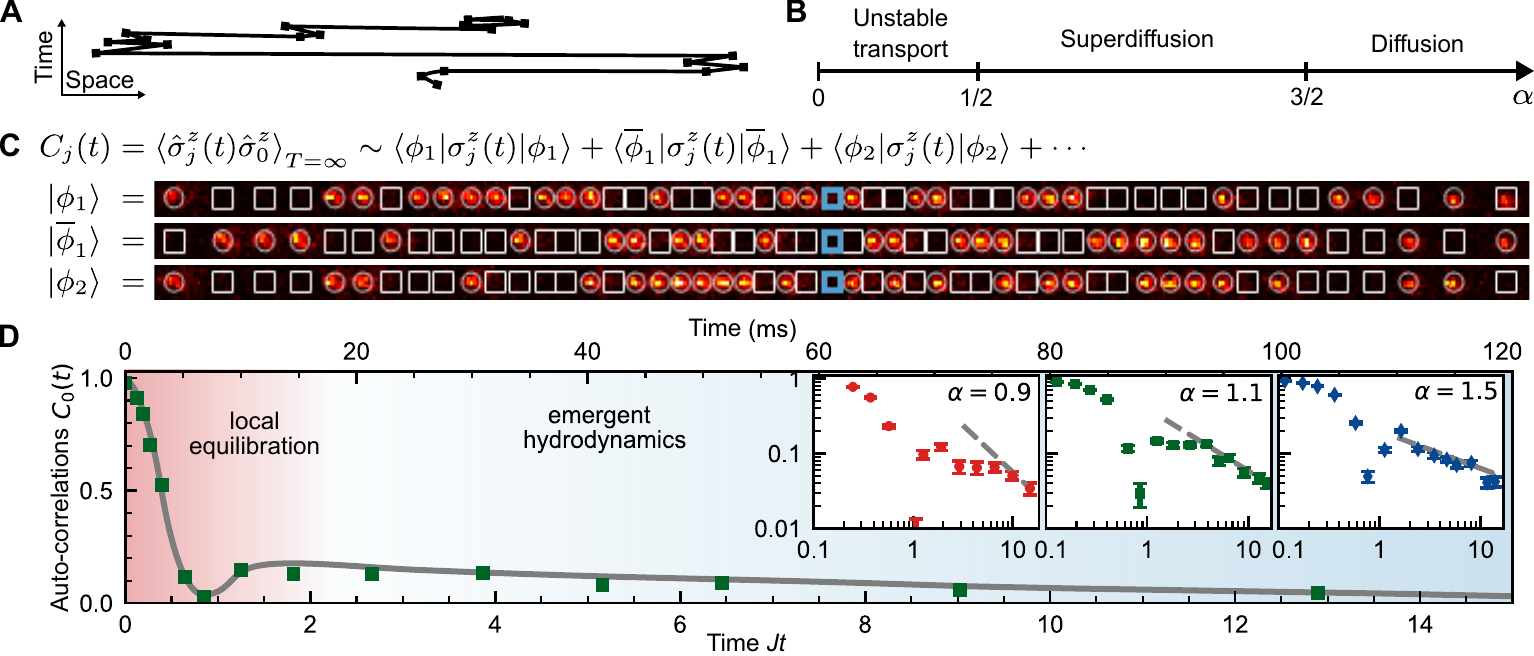}
\caption{\textbf{Emergent hydrodynamics in a long-range quantum magnet.} A) In our system, the dynamics of a locally created spin excitation is effectively governed by a hydrodynamic theory of L\'evy flights, which are random walks with step sizes drawn from a distribution with algebraic tails. This occasionally leads to long-distance jumps of the excitation. B) The family of hydrodynamic universality classes can be tuned by the power-law exponent $\alpha$ of the spin-spin interactions. C) We measure infinite-temperature correlations by averaging over initial product states while preparing the central ion deterministically in the same state (blue box); \cite{SI} Sec.~II. 
Picture of three exemplary initial states in a chain of $^{40}$Ca$^+$ ions ($\ket{\uparrow}$ dark spots, $\ket{\downarrow}$ bright spots). White squares (circles) indicate the intended preparation of $\ket{\uparrow}$ ($\ket{\downarrow}$), achieved with $99\%$ fidelity per ion. 
D) Measured auto-correlations (green squares) for 51 ions and $\alpha = 1.1$ (gray line: guide to the eye). 
Error bars, denoting the standard error of the mean  are smaller than the symbols (Sec.~V~\cite{SI}). 
At short times (red shading), the spin excitation quickly relaxes to a local equilibrium state. At late times (blue shading) global conservation laws constrain the relaxation of spin excitations, leading to a slow power-law decay of the auto-correlations. 
Insets: Auto-correlations on a double logarithmic scale for different values of $\alpha$ highlighting the tunable transport. Gray dashed lines are power laws with the predicted exponent from L\'evy flights. Here, $J = 248$ rad/s, $ 129$ rad/s, and $ 116$ rad/s for
$\alpha=0.9$ (51 ions), $\alpha = 1.1$ (51 ions), and $\alpha = 1.5$ (25 ions).
}\label{fig:auto}
\end{figure*} 
\begin{figure*}
\centering
\includegraphics{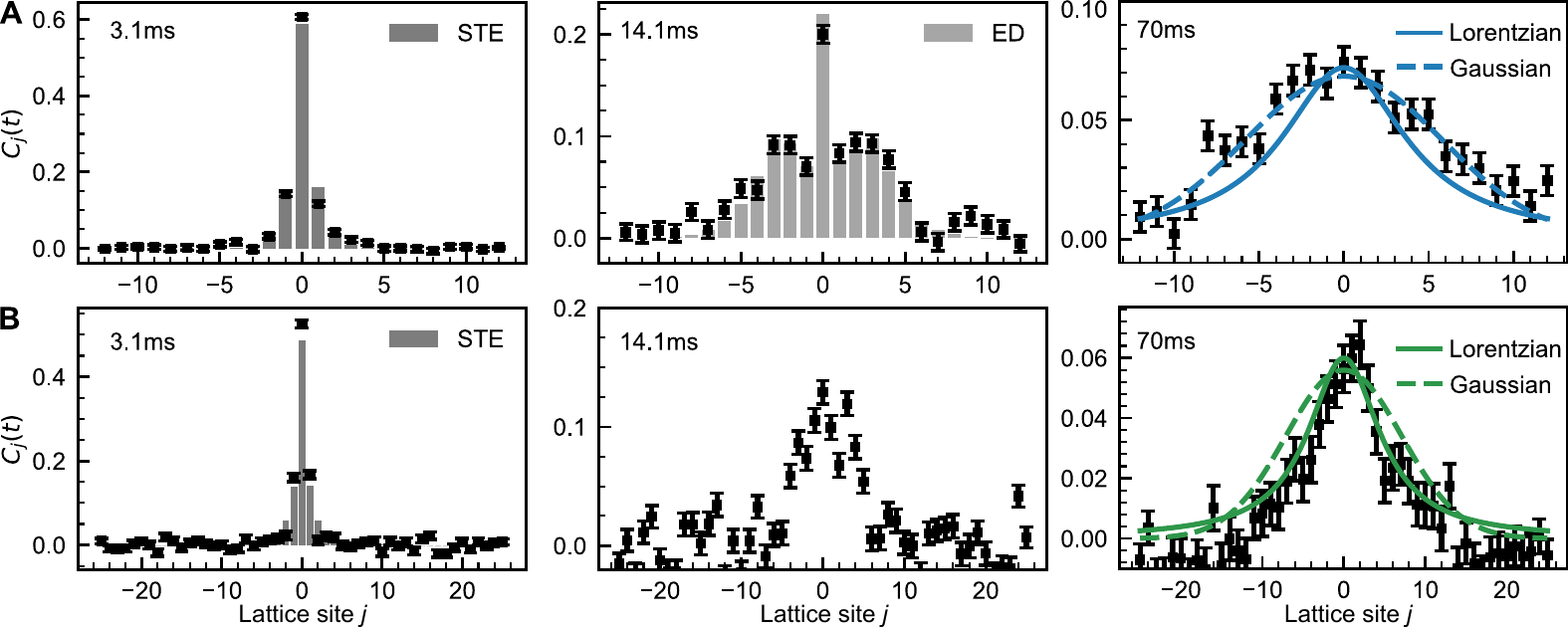}
\caption{\textbf{Spatial correlation profiles.} 
Spatially resolved correlations for A) $\alpha=1.5$ and B) $\alpha=1.1$. The spatial correlations are measured at short times (left), intermediate times (middle), and in the hydrodynamic late-time regime (right). STE: analytic short-time expansion, ED: exact diagonalization accessible only for the shorter chain of $25$ ions. In the hydrodynamic regime (right) the measured profiles are compared to predictions from L\'evy flights. The spatial profile in A) is compatible with a Gaussian (dashed) and in B) with a Lorentzian (solid), as supported by the reduced $\chi^2$ values of the fit: A) $\chi^2_{\mathrm{L}}=3.9$, $\chi^2_{\mathrm{G}}=1.3$; B) $\chi^2_{\mathrm{L}}=1.1$, $\chi^2_{\mathrm{G}}=3.6$ (obtained by fitting over the central 27 sites).
}\label{fig:space}
\end{figure*}

Conservation laws determine the  macroscopic late-time dynamics of quantum systems. In the long-range XY model the total magnetization $\sum_j \hat \sigma_j^z$ is conserved, which constrains the relaxation of spin excitations at late times. The long-range couplings enable transport of spin excitations over many lattice sites. As a consequence, the effective dynamics is expected to be governed by classical L\'evy flights comprised of long-distance jumps~\cite{PhysRevB.101.020416}, illustrated in Fig. \ref{fig:auto}A. These long-distance jumps become irrelevant when the variance of the step size of excitations is finite. In our model this is the case for $\alpha>3/2$, implying conventional diffusion. However, the variance diverges for $\alpha \leq 3/2$~\cite{RevModPhys.87.483} and hydrodynamics is expected to be described by the anomalous diffusion equation
\begin{equation}
    \partial_t \langle\sigma^z_j\rangle = D_\alpha \nabla^{2\alpha - 1} \langle{\sigma^z_j}\rangle,
    \label{eq:gendiff}
\end{equation}
with $D_\alpha$ denoting the associated transport coefficient; \cite{SI}, Sec. VIII.
 Therefore, transport is modified by the long-range character of the interactions, ranging from conventional diffusion over superdiffusion to a regime in which the spins are so strongly connected that the lattice geometry becomes irrelevant, and a mean-field regime is entered. The predicted dynamical phase diagram is shown in Fig. \ref{fig:auto}B. 

\begin{figure*}
\centering
\includegraphics{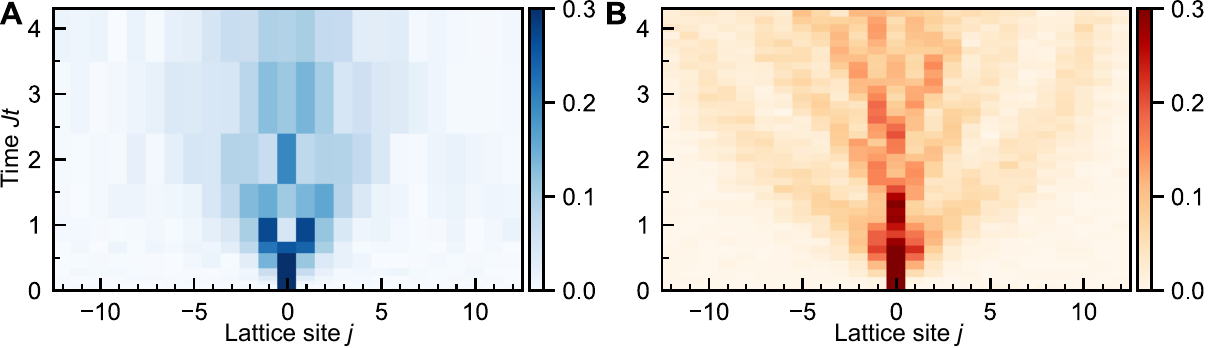}
\caption{\textbf{Contrasting the infinite temperature background with the spin polarized background.} A) The deterministically prepared excitation of the central ion strongly interacts with all the other excitations of the infinite temperature background and slowly spreads through the system following the laws of classical hydrodynamics. B) By contrast, a single excitation on top of the fully down-polarized state $\ket{\downarrow\cdots\downarrow\uparrow\downarrow \cdots\downarrow}$ has no other excitations to scatter off and therefore spreads freely, exhibiting quantum interference patterns. Data is measured for $25$ ions with power-law exponent $\alpha=1.5$. 
 }\label{fig:singlevsmultiple}
\end{figure*}

Hydrodynamic transport is most directly probed by creating a spin excitation at time $t=0$ in the center of the chain and tracking how it propagates in space and time, as measured by the unequal-time correlation function  
\begin{equation}
    C_j(t)=\langle{\hat\sigma^z_j(t)\hat\sigma^z_0}\rangle_{T=\infty}.
    \label{eq:def_Cj}
\end{equation}
Over the course of the quantum dynamics, the initial excitation has to scatter off a highly excited background in order to quickly reach the hydrodynamic regime. Therefore, we measure 
$ C_j(t)$ at infinite temperatures $T$ with magnetization $\sim0$. 

Both, measuring spin correlations at different times and preparing an infinite temperature state is difficult in isolated, engineered quantum systems. We overcome these challenges by expressing the infinite temperature expectation value as an equally weighted trace over product states $\ket{\phi}$ in the $\hat\sigma^z$ basis; Sec. X~\cite{SI}. When preparing the central spin in the same polarization, we have $\hat \sigma_0^z\ket{\phi} = + \ket{\phi}$; Fig. \ref{fig:auto}C. The correlation function can then be directly evaluated as $C_j(t) \sim \braket{\phi_1| \hat\sigma_j^z(t)| \phi_1}  +\braket{\phi_2| \hat\sigma_j^z(t)| \phi_2}  + \ldots$. It can be accurately obtained by sampling a finite number of initial product states; Sec. X~\cite{SI}. Yet, for a comparatively small and experimentally accessible number of initial states, large statistical fluctuations are expected in the measured correlations. We remove these fluctuations by sampling pairs of conjugate product states, $\ket{\phi}$ and $\ket{\overline{\phi}}$, where in the second configuration all spins are flipped except for the central one. For each pair of product states, initial correlations are unity in the center of the system while being zero elsewhere, reproducing directly this property of the full trace; see first two initial states in Fig. \ref{fig:auto}C. With this procedure, convergence is already achieved for a small number of initial product states; Sec. X~\cite{SI}. For $\alpha=0.9$ and $1.1$ ($\alpha =1.5$), we create 60 (120) initial product state configurations each of which is realized, evolved, and measured for 50 to 200 times.

The auto-correlation function $C_0(t)$  determines the residual excitation in the center of the chain. We measure $C_0(t)$ for 51 ions; 
Fig. \ref{fig:auto}D. The relaxation dynamics occurs in multiple stages. A local equilibrium is reached already after a few collisions with the abundant excitations of the infinite temperature state. As a consequence, at short times the auto-correlation exhibits a rapidly damped oscillation. At later times the system enters the hydrodynamic regime and eventually approaches a global equilibrium through the slow rearrangement of spin excitations constrained by the conserved magnetization. The power-law decay of the correlations becomes manifest
on a double logarithmic scale; see insets for various values of $\alpha$. Our experimental data is consistent with the hydrodynamic theory of Eq.~\eqref{eq:gendiff}, which predicts a power-law exponent of $1/(2\alpha-1)$. Exploiting the 
experimental control over the long-range exponent $\alpha$, hence enables us to tune the transport from normal diffusion to anomalous superdiffusion.

\begin{figure*}
\centering
\includegraphics{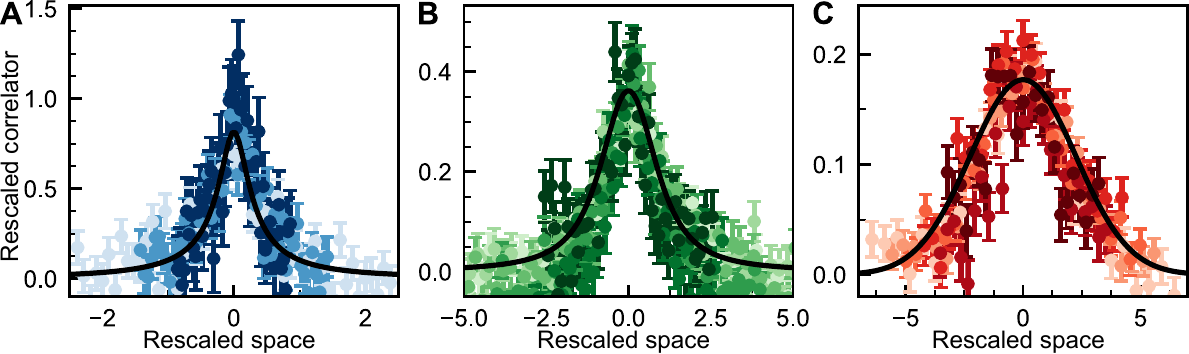}
\caption{\textbf{Extracting transport coefficients.} 
At late times the correlations exhibit a self-similar scaling, relating space and time with a microscopic transport coefficient $D_\alpha$. Rescaled correlations $\tau_x C_j(t)$ as a function of rescaled space $j/\tau_x$, where $\tau_x=(Jt)^{1/(2\alpha-1)}$. A) $\alpha = 0.9$, 51 ions, B) $\alpha = 1.1$, 51 ions, and C) $\alpha = 1.5$, 25 ions.  Correlations are shown for times $Jt>5$  (darker colors correspond to later times).}\label{fig:collapse}
\end{figure*}

A more stringent test of the emergent hydrodynamics is obtained from the full spatial correlation profile; Fig \ref{fig:space}. We realize the long-range spin model for $\alpha = 1.5$ with 25 ions and for $\alpha = 1.1$ with 51 ions. At early times (3.1ms) the quantum dynamics is well-described by an analytic short-time expansion of the equations of motion (\cite{SI} Sec. IX). At intermediate times (14.1ms), the excitation starts spreading through the system, but some quantum coherence remains, indicated by the spatial oscillations. For 25 ions, the measured dynamics compares well with results obtained from exact diagonalization, demonstrating the coherence of our quantum system; Sec VII \cite{SI}. For the 51 ion chain a comparison to exact diagonalization is not possible due to the exponential growth of the Hilbert space, reaching a dimension of $2^{51}\approx 10^{15}$.

At the latest times shown, interactions have entangled local degrees of freedom with the rest of the system. Quantum interference patterns are averaged out and the hydrodynamic regime is entered. This becomes even more apparent in the space-time correlations; Fig. \ref{fig:singlevsmultiple}A. By contrast, a single excitation on top of a spin-polarized state cannot scatter, and the associated correlations exhibit coherent space-time oscillation patterns instead~\cite{jurcevic_quasiparticle_2014}; Fig. \ref{fig:singlevsmultiple}B.

The theory of L\'evy flights predicts the following scaling form for the spatio-temporal profiles 
\begin{equation}
    C_j(t) = (D_\alpha t)^{-\frac{1}{2\alpha -1}} F_\alpha\left(\frac{|j|}{(D_\alpha t)^{1/(2\alpha-1)}}\right),
\label{eq:scaling_form}
\end{equation}
where $F_\alpha$ is given by the family of stable symmetric distributions~\cite{RevModPhys.87.483}. The scaling between space and time entering the distribution $F_\alpha$ can be deduced directly from the generalized hydrodynamic equation \eqref{eq:gendiff}. The distribution $F_\alpha$ cannot be expressed in terms of elementary functions, except for $\alpha = 3/2$, where a Gaussian indicates conventional diffusion, and $\alpha = 1$, where a Lorentzian appears. 

The measured hydrodynamic profiles are compatible with a Lorentzian for $\alpha=1.1$  ($F_{\alpha=1.1}$ is still very close to a Lorentzian) and with a Gaussian for $\alpha=1.5$, see $\chi^2$ analysis in Fig. \ref{fig:space} (right panels), in agreement with Eq.~\ref{eq:scaling_form}. 

One of the most striking predictions of hydrodynamics is the self-similar scaling of the correlations, relating time and space by a universal dynamical scaling exponent, given by $1/(2\alpha-1)$, and a non-universal constant, the transport coefficient $D_\alpha$. In analogy with the universal scaling of correlation functions near a second-order equilibrium phase transition, the hydrodynamic scaling indicates the proximity of the dynamics to a thermal fixed point. While the scaling form of the data is universal and can be predicted from purely hydrodynamic reasoning, the transport coefficient $D_\alpha$ depends on the full quantum many-body spectrum and is therefore challenging to predict from analytical or numerical methods. From a scaling collapse of the experimental data, shown in Fig. \ref{fig:collapse}, we obtain a transport coefficient of  $D_\alpha/J=0.5_{-0.1}^{+0.2}, 0.8_{-0.2}^{+0.3}, 2.6_{-0.7}^{+0.9}$ for $\alpha=0.9, 1.1, 1.5$~(Sec. VI \cite{SI}).

Large, coherent quantum systems with local control provide key insights into fundamental properties of non-equilibrium quantum states. In our experiment, we measure the emergent macroscopic hydrodynamics in an infinite temperature state of a strongly interacting spin chain; a regime that is notoriously challenging to describe by controlled analytical or numerical calculations.
By measuring the full spatio-temporal profile of the hydrodynamic scaling functions, we experimentally establish a tunable family of transport ranging from conventional diffusion to anomalous superdiffusion. 

An exciting prospect is to investigate transport in models with enhanced symmetries, which can be realized in our setting by suitably designed Floquet protocols~\cite{birnkammer_2020}. Moreover, our tools for measuring high-temperature correlations can be readily applied to other quantum devices with local control, including quantum gas microscopes, Rydberg atom arrays, and superconducting qubits. New classes of hydrodynamics can be realized and probed in that way.

\textit{Note added}: During the completion of this manuscript, we became aware of related work demonstrating superdiffusive transport in an integrable Heisenberg chain with nearest-neighbor superexchange interactions~\cite{Bloch}. 

\textbf{Acknowledgements}. 
We acknowledge support from the Technical University of Munich - Institute for Advanced Study, funded by the German Excellence Initiative and the European Union FP7 under grant agreement 291763, the Max Planck Gesellschaft (MPG) through the International Max Planck Research School for Quantum Science and Technology (IMPRS-QST), the Deutsche Forschungsgemeinschaft (DFG, German Research Foundation) under Germany’s Excellence Strategy--EXC--2111--390814868. The project leading to this application has received funding from the European Union’s Horizon 2020
research and innovation programme under grant agreement No 817482, from the European Research Council (ERC) under the European Union’s Horizon 2020 research and innovation programme (grant agreement No. 851161, No. 741541 and No. 771537), the Institut f\"ur Quanteninformation GmbH, and Austrian Science Fund through the SFB BeyondC (F7110).

\textbf{Author contributions}. The research was devised by MJ, CM, MK, and CFR. AS, IL and MK developed the theoretical protocols. MJ, FK, CM, RB, and CFR contributed to the experimental setup. MJ and FK performed the experiments. MJ, AS, IL, MK and CFR analyzed the data and MJ, AS, and CFR carried out numerical simulations. MJ, AS, MK, and CFR wrote the manuscript. All authors contributed to the discussion of the results and the manuscript. All correspondence should be addressed to MK (michael.knap@ph.tum.de) and CFR (christian.roos@uibk.ac.at)

\textbf{Competing interests.} The authors declare no competing interests.
 \newpage \ \newpage
\section*{\large \textbf{Supplemental materials}}

\section{Experimental setup}
The experiments are performed with a string of $^{40}$Ca$^+$ ions confined in a linear radiofrequency trap. For confining long ion strings, the trapping potential is made strongly anisotropic, with oscillation frequencies of 2.93 MHz and 2.898 MHz in the radial plane, and an axial trapping frequency of 126.3 kHz (126.6 kHz) for trapping 25 ions (51 ions). At the above stated confining frequencies, the 25-ion (51-ion) string has a length of 173 $\mu $m (247 $\mu$m). (Pseudo-)spins are encoded in superpositions of 4S$_{1/2},m=1/2$ and 3D$_{5/2},m=5/2$ electronic states and coherently manipulated by a frequency-stable laser at 729~nm coupling to the quadrupole transition. A low-noise magnetic field of 4.17~Gauss produced by an assembly of SmCo magnets is used to lift the degeneracy of the Zeeman states. 

At the beginning of each experiment, all transverse collective modes of the ion string are cooled close to the ground state by sideband cooling on the S${}_{1/2}\leftrightarrow \text{D}{}_{5/2}$ quadrupole transition. The axial modes are polarization-gradient cooled to sub-Doppler temperatures and the ions are prepared in the S$_{1/2},m=+1/2$ state by frequency-resolved optical pumping \cite{Joshi_2020}.

At the end of each experimental run, a quantum state measurement projecting onto the (pseudo-)spin basis states is carried out by spatially resolved measurements of the ions' fluorescence emitted on the S$_{1/2}$ to P$_{1/2}$ transition. The state assignment error probability is below $10^{-3}$ per ion.

\begin{figure*}
\centering
\includegraphics{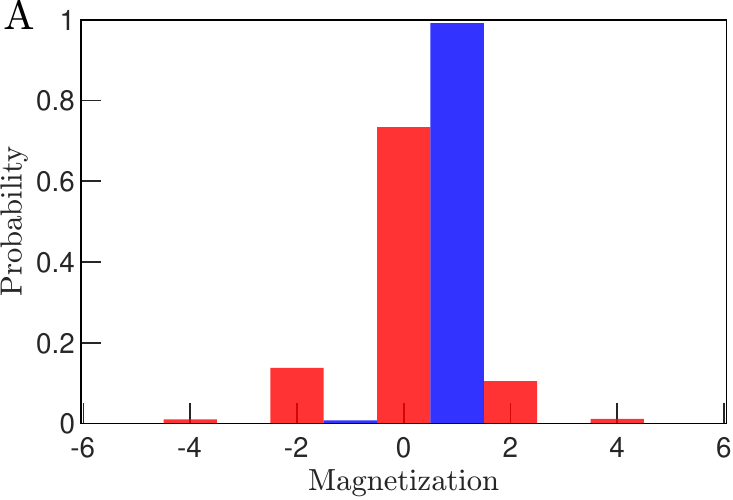}
\includegraphics{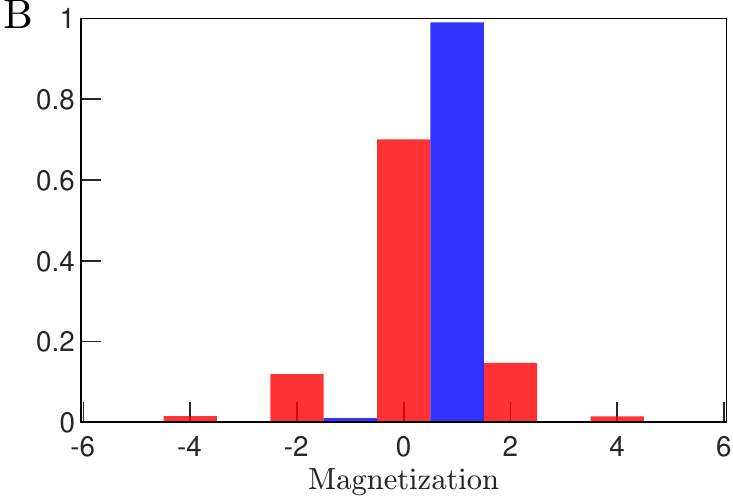}
\caption{Initial state preparation fidelities of the A) 25-ion and B) 51-ion strings. Individual ions are prepared in spin up/down states with about $99\%$ fidelity. Blue bars indicate the magnetization of the center ion whereas red bars indicate the total magnetization of all other ions.} \label{fig:InitialStatePrep}
\end{figure*}

\section{Initial state preparation}
Preparation of the (pseudo-)spins in the desired initial product states is achieved by using a spin-echo sequence that combines resonant laser pulses coupling to all spins with nearly the same strength and off-resonant pulses addressed to individual spins that are sandwiched between the global pulses. For a 25-ion crystal, the state preparation starts with a set of composite $\pi/2$ pulses that prepares all calcium ions in an equal superposition of the 4S$_{1/2},m=1/2$ and 3D$_{5/2},m=5/2$ electronic states. A tightly focused laser beam, with beam radius of about 2 $\mu$m, is steered over the ion string (ion spacing $>$ 4 $\mu$m) with an acousto-optic deflector. Pre-calibrated addressing pulses are placed in a specific order within the two halves of the spin-echo sequence such that errors resulting from cross-talk while addressing neighbouring ions are minimized. The pulse sequence is completed by rotating the unaddressed (addressed) ions to the S (D) state by the second global $\pi/2$ pulse. For a 51-ion crystal, we perform the above described sequence but on 4S$_{1/2},m=1/2$ and 3D$_{5/2},m=-3/2$ electronic states before restoring the population to the desired qubit states. Due to the restriction of the polarization angle and beam geometry, we only achieve optimum off-resonant coupling, thus short interaction times, with the addressing beam while working on 4S$_{1/2},m=1/2$ and 3D$_{5/2},m=-3/2$ states.

\textbf{Composite laser pulses:} Ions are collectively excited with an elliptically shaped laser beam with a beam diameter of about 470 $\mu$m, which is shone over the ion string of length 247 $\mu$m from the transverse direction of the ion string. This leads to a non-uniform illumination, resulting in spatially dependent rotation angles of the ions when performing the global single qubit rotations. In our experiment, we reduce the impact of these unequal rotations of the spins by employing Composite pulse sequences \cite{Torosov2012}. The Composite pulse sequences for $\theta=\pi/2$ and $\theta=\pi$ angle rotations are given as
\begin{eqnarray}
    U^{x}_{\theta=\pi/2}=e^{-i\frac{\pi}{4}\sigma_y}e^{-i\frac{\pi}{4}\sigma_x},\\
    U^{x}_{\theta=\pi}=e^{-i\frac{\pi}{4}\sigma_x}e^{-i\frac{\pi}{2}\sigma_y}e^{-i\frac{\pi}{4}\sigma_x}.
\end{eqnarray}

\textbf{Off-resonant rotations with a tightly focused beam:} The tightly focused off-resonant laser beam is used in conjunction with the global laser beam. Pulse length and laser intensity of the addressing beam are chosen such that the specific ions are prepared in the desired quantum states with a single ion fidelity of about $99\%$. Addressing of 51(25) ions, required about 0.5(0.25) ms of interaction time, which is much shorter than the coherence time ($\sim$ 50 ms) of our system, hence the initial state is well protected against any decoherence effects. Slow spatial drift of the ion string along the weak confining axis, which impacts the quality  of addressing operations, is corrected on-the-fly by routinely measuring the ion positions and consecutively applying compensating potentials to the trap electrodes. Laser beam intensities of both global and addressing laser beams are stabilized via a sample and hold intensity stabilization technique. Furthermore, the addressing pulse lengths are routinely calibrated and corrected against any unwanted slow variations. 

\textbf{Product state fidelities:} Fig. \ref{fig:InitialStatePrep} shows initial state preparation fidelities of the 25-ion and 51-ion strings. Each spin is prepared in the desired up/down basis states with about 98.8$\% (99.3\%)$ fidelity, which corresponds to  a fidelity of about $ 73$\%$(70\%)$ of the desired product state for a 25(51) ion string.

\begin{figure*}
\centering
\includegraphics{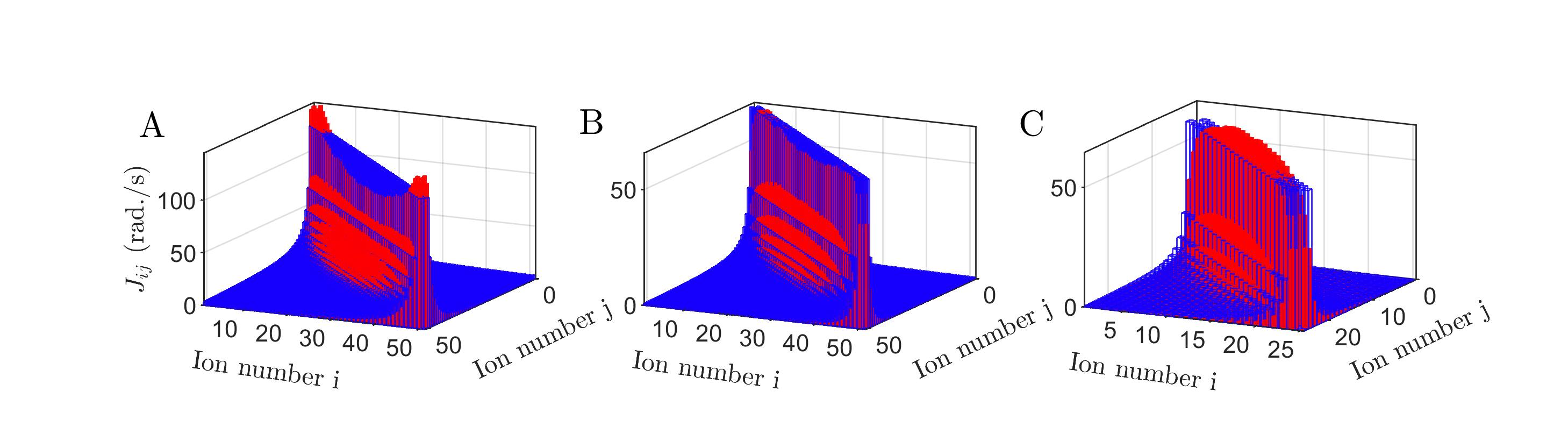}
\caption{Spin-spin interaction matrix terms displayed for A) $\alpha =0.9$, B) $\alpha =1.1$, and C) $\alpha =1.5$. Bars in red represent experimentally realized interaction elements and bars in blue represent the interaction elements that are approximated by a power law.} \label{fig:ApproxPowLaw}
\end{figure*}

\begin{figure*}
\centering
\includegraphics{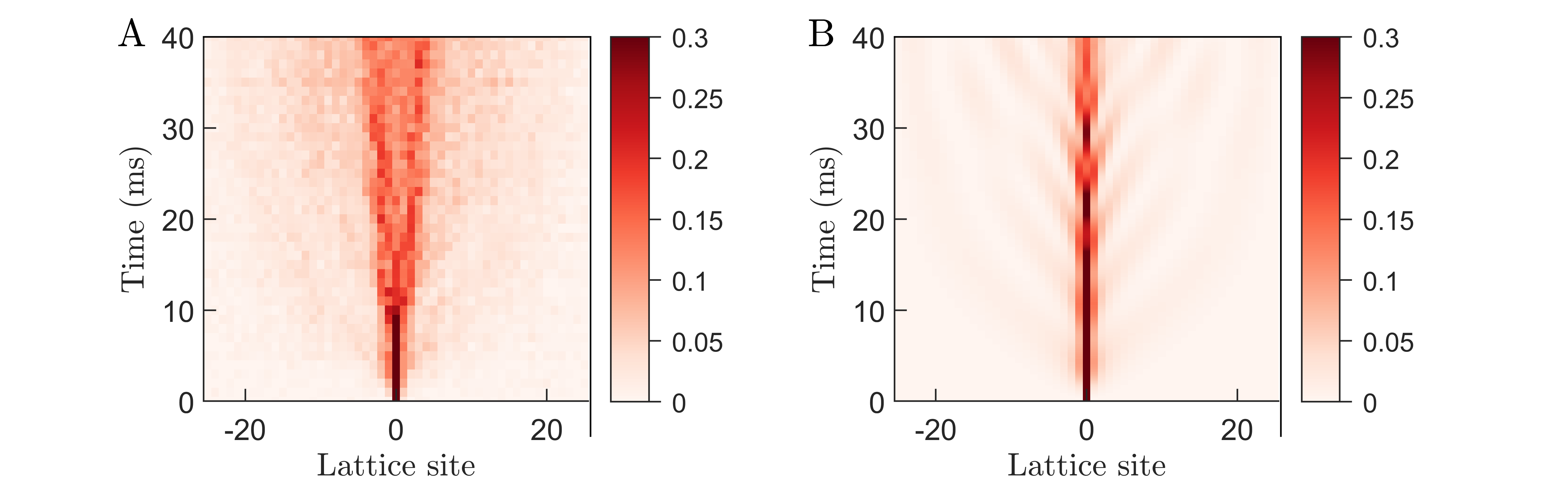}
\caption{A) Experimental spread of excitation as a function of entangling laser-ion interaction time for $\alpha =1.1$ (51 ions). Here we initialize the middle ion as spin-up and remaining ions as spin-down. B) Numerical simulations performed for the same experimental conditions.} \label{fig:SingleExcWalk}
\end{figure*}

\section{Engineered spin-spin interaction and many body dynamics }
Spin-spin interactions are engineered with a two-tone laser beam inducing M{\o}lmer-S{\o}rensen type interactions; i.e. the laser off-resonantly couples to both the lower and upper vibrational sidebands of all $2N$ collective motional modes describing the ion motion in the plane that is normal to the direction of an $N$-ion string. The tunable long-range coupling is provided by a specific choice of laser parameters that generates a long-range Ising-type spin-spin interaction with a coupling strength approximately described by a power-law~\cite{Porras_2004, Nevado_2016}. The laser beam is elliptically shaped and elongated along the axis of the ion string to provide near-uniform illumination over the ion string. Unwanted light shifts, which occur due to the presence of other electronic transitions, are compensated with a third laser beam that is mixed with the two-tone laser beam \cite{jurcevic_quasiparticle_2014}. The laser beam propagates from the transverse direction of the axis of the ion string and imparts a near flat waveform over the ion string, which otherwise can have adverse effects in the laser-ion interaction. 

The engineered spin-spin interaction, for $B \gg J$ gives rise to a flip-flop type interaction described by  
\begin{equation}\label{eq:powerlawexp}
    \hat H= \sum_{i<j} \frac{J}{|i-j|^\alpha} (\hat \sigma^+_i\hat\sigma^-_j + \hat \sigma^-_i\hat\sigma^+_j),
\end{equation}
where exponent $\alpha$ and strength of the spin-spin interaction $J$ can be tuned by changing the detuning and intensity of the entangling laser beam. The experimentally realized Hamiltonian slightly differs from the power law expression given in equation \ref{eq:powerlawexp}. The exact spin-spin matrix connecting all spins is given by
\begin{equation}
    J_{ij}=\frac{\Omega_i \Omega_j}{2}\sum_{m=1}^{2N}\frac{\eta_{im}\eta_{jm}}{\Delta_m},
\end{equation}
where $\eta_{jm}$ is the Lamb-Dicke parameter of the $j^{th}$ ion and the $m^{th}$ motional mode. $\Delta_m$ is the detuning of the two-tone laser field from the motional mode of interest and $\Omega_i$ is the Rabi frequency of the $i^{th}$ ion. 
The detuning of the laser beams from the vibrational sideband ranges from a few tens of kHz for the center-of-mass (COM) mode to about 1.5 MHz for the lowest-frequency mode, resulting in effective spin-spin interactions that decay with distance between the spins. The structure of the coupling matrix can be experimentally controlled by varying the laser detuning from the COM mode and the anisotropy of the trapping potential.
A comparison between the approximated power law and experimentally realized interaction Hamiltonian is shown in Fig. \ref{fig:ApproxPowLaw} for all three $\alpha$ values that are used in the main manuscript. The figure displays a small mismatch between the power-law interaction of eq.~\ref{eq:powerlawexp} and the effective interaction realized in the trapped ion systems \cite{Nevado_2016}, which is inherently present in any the laser-ion interaction. Additionally, due to technical reasons, a non-uniform illumination of the laser beam  over the ion chain alters the spin-spin interaction and slightly weakens the interactions of the spins at the ends of the chain. The numerical simulations presented in the main manuscript are performed for the experimentally realized spin-spin interaction which includes the non-uniform laser intensity over the ion string.

The engineered entangling interaction is experimentally tested by time evolving an initial state where the middle ion (remaining ions) is initialized in a spin-up (spin-down) state. In Fig. \ref{fig:SingleExcWalk}(a) we show dynamics of spin excitation for $\alpha =1.1$. We compare the experimental results with numerical simulations that we carry out in a subspace where the magnetization is conserved, see Fig. \ref{fig:SingleExcWalk}(b) and find a good agreement at early times whereas some discrepancy is visible at late times. This discrepancy is attributed to decoherence effects that arise from non-perfect initial state preparation and non-conservation of magnetisation, discussed in the following section. 

\section{Decoherence effects}

While experimentally realizing the spin-spin Hamiltonian, the system undergoes some unwanted decoherence effects that can alter the dynamics. In this section, we describe the two major decoherence channels that are important for the initial state preparation and dynamics with long ion strings. The first one is dephasing noise, which arises from laser frequency/phase noise and fluctuating magnetic fields. Jointly, they reduce the coherence time of our spins. Most importantly, the superposition, which is an essential step for individual control of the spins for preparing arbitrary product states, decays due to the dephasing noise. The laser frequency noise is reduced by locking the laser to an ultra-stable reference cavity. The trap setup is placed inside a mu-metal shield to suppress the magnetic field noise. The overall coherence time of the system is about 50 ms. Furthermore, we reduce the impact of the dephasing noise by applying spin-echo sequences. While these effects play an important role for the state preparation, the dynamics is not affected, as we work in the decoherence-free-subspace in which the spins are protected against the dephasing noise.

\begin{figure}
\centering
\includegraphics[width=\columnwidth]{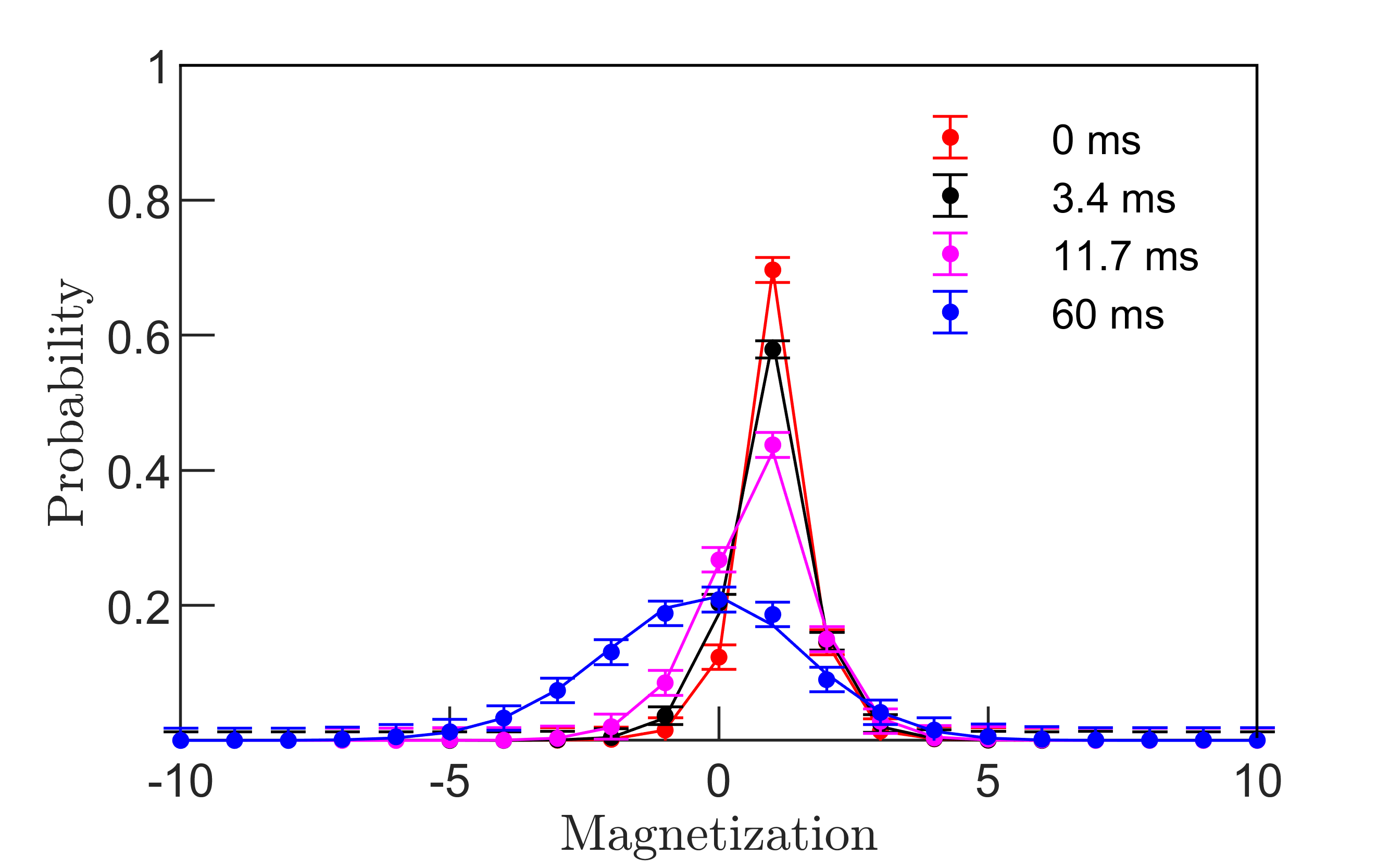}
\caption{Magnetization variation as a function of interaction time. Filled circles are experimental data points and solid lines are fitted to the experimental data. Here we randomly sample the initial states while keeping the total magnetization to be +1. The measurement protocol is identical to the one we described in Fig. 1 of the main manuscript. Error bars are estimated to be 1/$\sqrt{N_m}$, where $N_m$ is number of measurements.} \label{fig:MagVar}
\end{figure}

The second type of decoherence that spins experience can be described by amplitude damping and depolarization channels. Due to the finite lifetime of the metastable state D${}_{5/2}$ ($\tau \sim $1 s), in which our spin-up state is encoded, decays to the S${}_{1/2}$ electronic state. In addition to this decay channel, the spins experience unwanted excitation/decay due to the bichromatic laser field coupling to the motional sidebands. These processes violate the conservation of magnetization that would be expected from the flip-flop interaction of eq.~(\ref{eq:powerlawexp}). We experimentally measure the spin-flip rates for the experimental conditions that are used in the main manuscript. In Fig. \ref{fig:MagVar} we show magnetization variation as a function of interaction time for a 51-ion chain. Here, we model two spin flip rates: spontaneous decay of the excited state $\Gamma$ and the laser induced spin flips rate $\gamma_\text{flip}$, similar to what has been described in reference \cite{Brydges_2019}. The model assumes that the spin-flip rate due to incoherent laser-ion coupling does not depend on the spin orientations. The flip rates that we extract by fitting the experimental data are $\gamma_\text{flip} =0.78$/s and $\Gamma=0.91$/s. These two competing processes in the end give rise to a dominating decay of the spins that are encoded as spin up over those are encoded as spin down. For the measurements that are presented in this manuscript, we remove the biased decay of the excitation to leading order by preparing a second set of initial states, where the central spin is prepared in a $\ket{\downarrow}$-state and average over the two distinct sets of initial states.  

\begin{figure}
\centering
\includegraphics[width=\columnwidth]{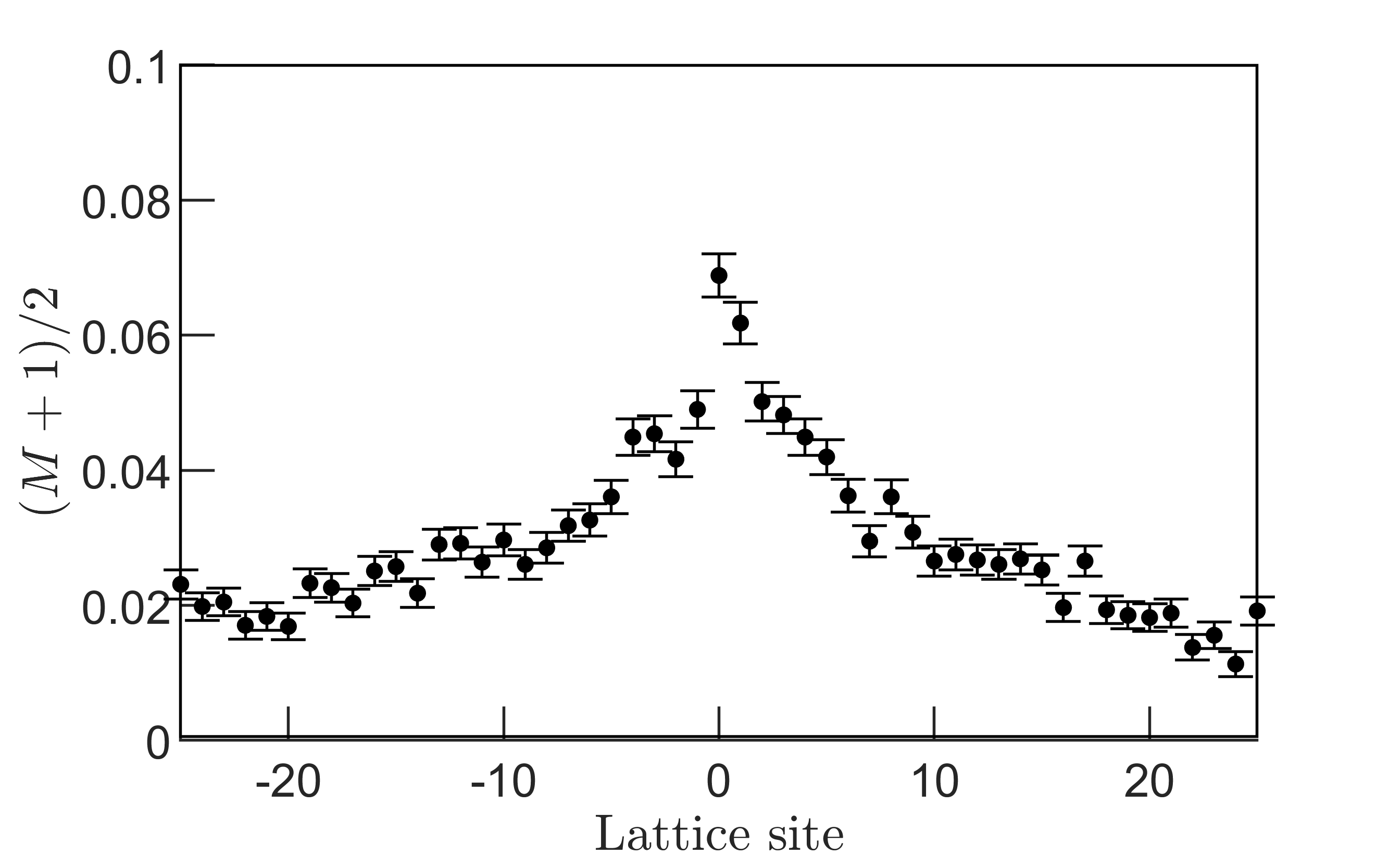}
\caption{Position dependent variation of the magnetization of a 51-ion chain after 40 ms of entangling operation. Here, all the ions are prepared in a spin-down product state before switching on the entangling interaction. } \label{fig:MagVariaionSpatially}
\end{figure}

\begin{figure}
\centering
\includegraphics[width=\columnwidth]{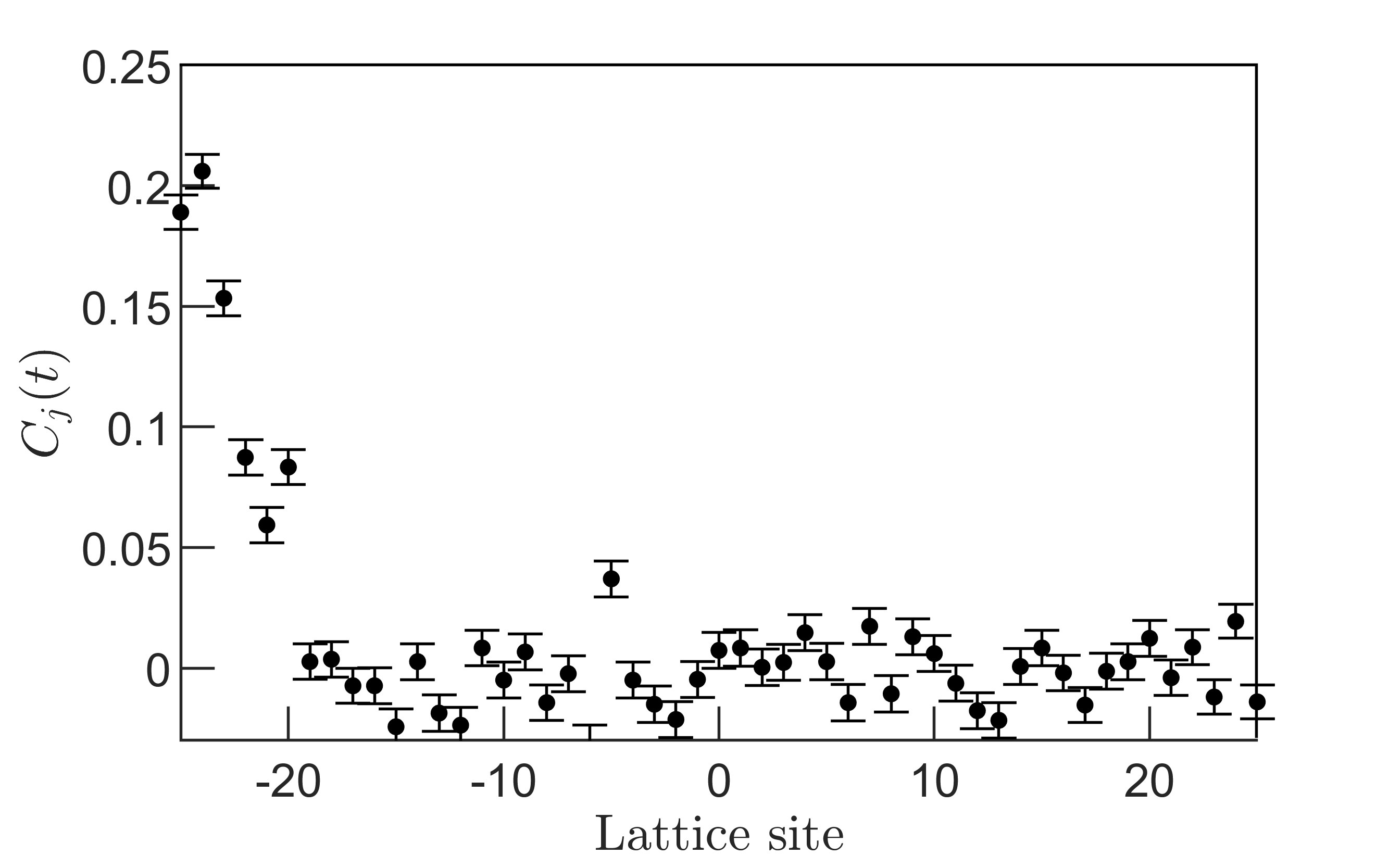}
\caption{Correlation measurements shown for 51 ions after 40 ms of entangling operation, while storing the excitation at one of the edges of the ion string. The experimental sequence is identical to that has been discussed in the main text. } \label{fig:HydroAtIon1}
\end{figure}

The magnetization decay model, which we assumed for fitting of the experimental data, considers that all the ions have the same spin flip rate. However, we experimentally observed that this rate depends slightly on the position of the ions in the string. The exact mechanism underpinning this effect is still under investigation. We speculate that the spatially dependent laser-ion interactions, coupling to the low-frequency motional modes of the ion chain, could induce such adverse effects. Most importantly, cross-frequency components that arise while producing the entangling laser beam could excite motional modes which in an ideal scenario should be suppressed. Fig. \ref{fig:MagVariaionSpatially} displays one of these artifacts which is observed while time evolving a spin state with all spins prepared in spin-down orientation. It may be noted that the magnetization varies the most for the middle ions in comparison to the outer ions. Ideally, the Hamiltonian does not couple this state to any other states hence we expect the state to be unaltered. However, due to the incoherent spin flips the total magnetization of the time evolved state changes with the interaction time. We further investigate the impact of spatially dependent incoherent processes in our current study of emergent hydrodynamics. In Fig. \ref{fig:HydroAtIon1}, we show the correlation function that we experimentally measure at $t=40$ ms. The measurement sequence is identical to that has been described in Fig. 1 of the main manuscript, except that we exchange the encoding of ion number 26$^{\text{th}}$ with the first ion in the 51-ion string. Note that the current measurements reveal no signature of unwanted variation of the magnetization which signifies the robustness of the measurement protocol against magnetization errors.

\begin{figure*}
\centering
\includegraphics[width=57mm]{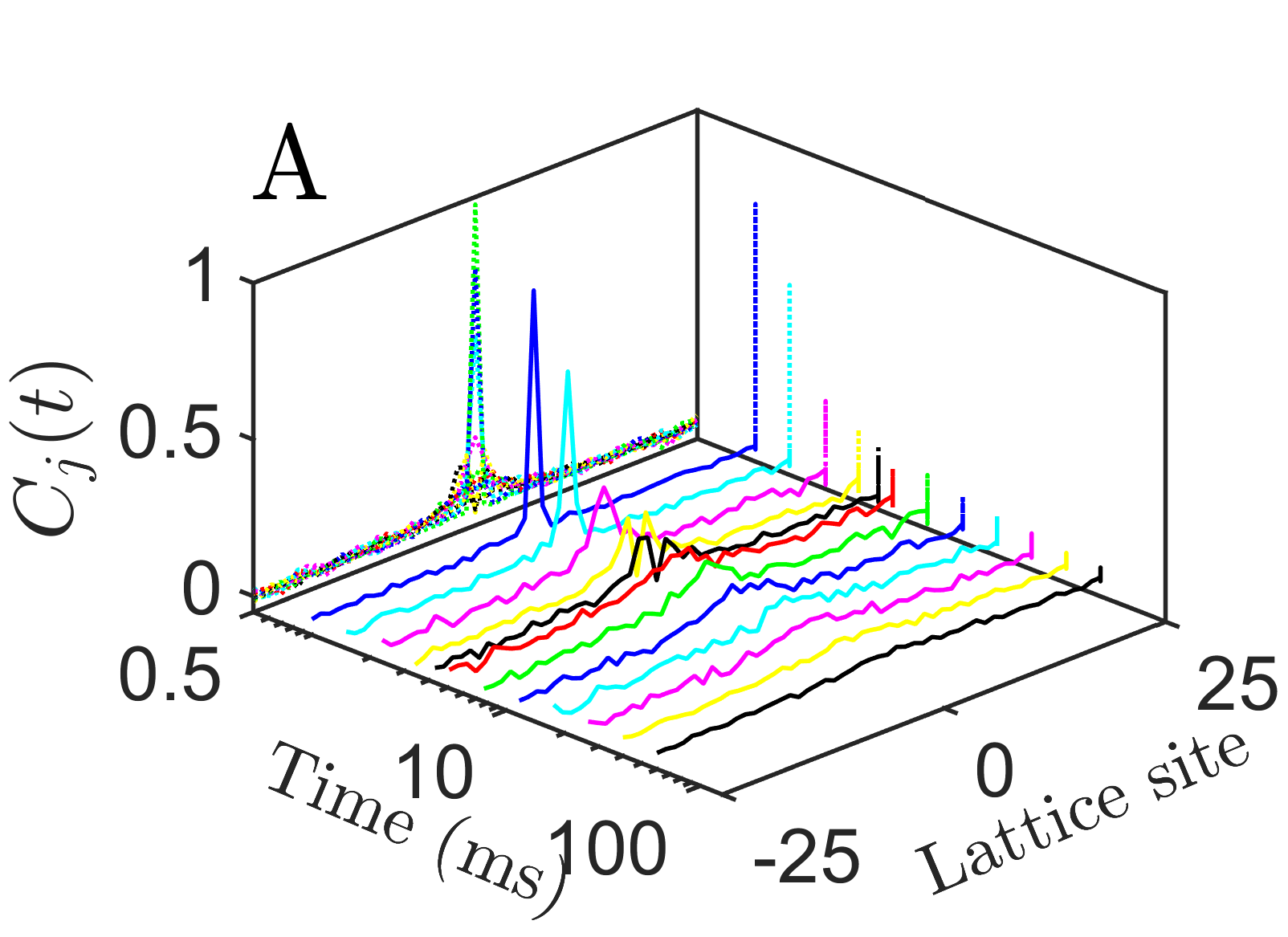}
\includegraphics[width=57mm]{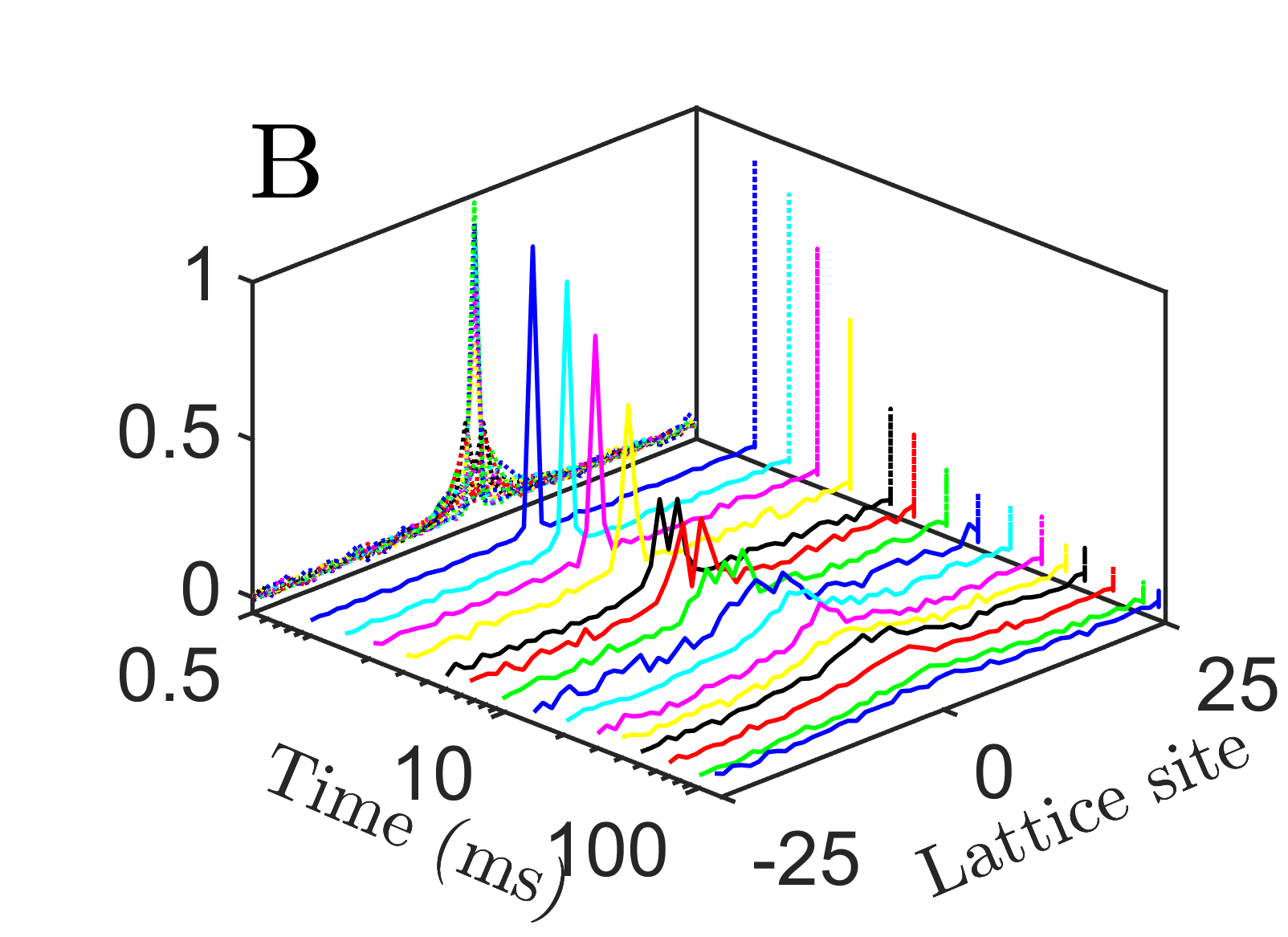}
\includegraphics[width=57mm]{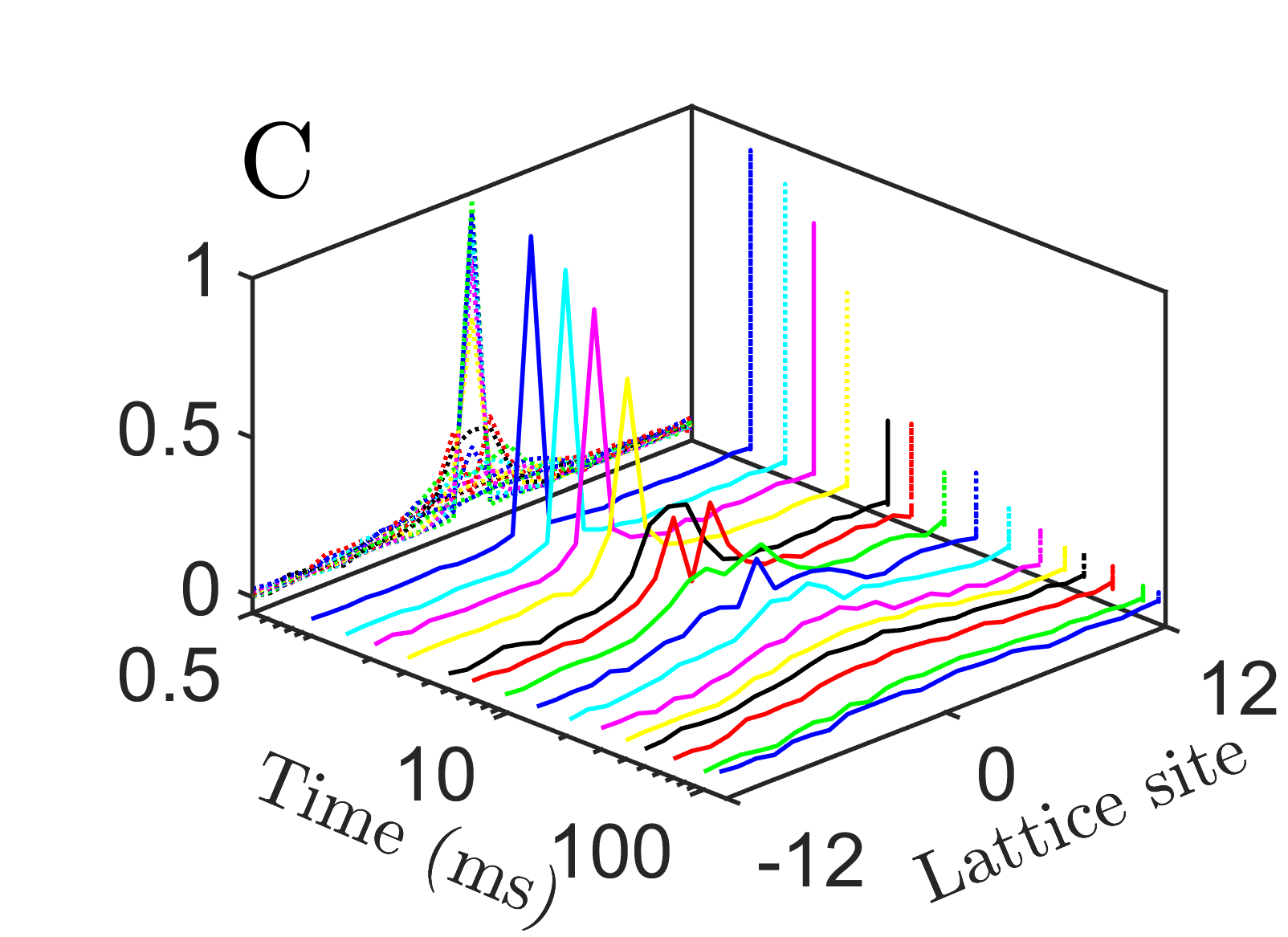}
\caption{Spatio-temporal evolution of the correlations for A)  $\alpha=0.9$ (51 ions), B) $\alpha=1.1$ (51 ions) and C)  $\alpha=1.5$ (25 ions). Dashed lines represent projection of each curve in the `XZ' and `YZ' planes. }  \label{fig:UnScSpatioTempo}
\end{figure*}

\section{Error estimation}
For the present measurements, the experimental error bars for an individual ion are estimated from the quantum projection noise model \cite{Itano1993}. In our experiment, we repeat the measurements $N_m$ times for each initial state. Each initial state is chosen from a uniform random distribution function that samples the individual spin state with an equal probability of spin-ups and spin-downs, while restricting the total magnetization of the whole system to be $+1$. The measurements are performed $N_u$ times. The mean standard deviation of  measurements for $i^\text{th}$ ion, while preparing the initial state $\ket{\Phi_u}= \bigotimes_{i = 1}^{N} \ket{\phi_u^i}$, is given by
\begin{equation}
    \sigma_u^{i}=\sqrt{p_u^{i}(1-p_u^{i})/N_m},
\end{equation}
where $p_u^{i}$ is the probability of finding the $i^{th}$ ion, prepared in $\ket{\phi_u^i}$  state, in the measurement basis. We further propagate this formula for the auto-correlation function described in Eq. (3) of the main text and estimate the mean standard deviation to be
\begin{equation}
    \sigma_{corr}=2\frac{\sqrt{\sum_{u=1}^{N_u} \sigma_u^2}}{N_u}.
\end{equation}.

\section{Transport coefficients} To extract transport coefficients from the experimental data, we perform a least-square fitting with the spatio-temporal function
\begin{equation}
    C_j(t)=(D_\alpha t)^{-\beta} F_{\alpha}\left(\frac{|j|}{(D_{\alpha}t)^\beta}\right),
    \label{eq:scalingansatz}
\end{equation}
where $D_\alpha$ represents the transport coefficient and the exponent $\beta$ determines the rate at which auto-correlations decay with time.  $F_\alpha$ are the stable symmetric distributions defined in Eq.~\ref{eq:F_alpha} below.

\textbf{Fitting of scaling functions:} In Fig. 2 of the main text we fitted a normalized Lorentzian
and Gaussian
to the data points which are finite within error bars. Hence, the full profile was fitted for $\alpha=1.5$, while the fit was restricted to the central 27 sites for $\alpha=1.1$.

\textbf{Extraction of diffusion coefficients: } In Fig. 4 of the main text, we used the expected values for the exponent $\beta=1/(2\alpha-1)$ to rescale the data. We then fitted the Ansatz in Eq.~\ref{eq:scalingansatz} to the rescaled data to extract the diffusion coefficient $D_\alpha$. The error bars are deduced by taking into account the uncertainty in extracting the $\alpha$ parameters.  

\textbf{Spatio-temporal evolution of correlations: }
The unscaled spatio-temporal evolution of correlations for all three $\alpha$ values are shown in Fig.~\ref{fig:UnScSpatioTempo}. 

\section{Auto-correlation function}
\begin{figure}
    \centering
    \includegraphics[width=\columnwidth]{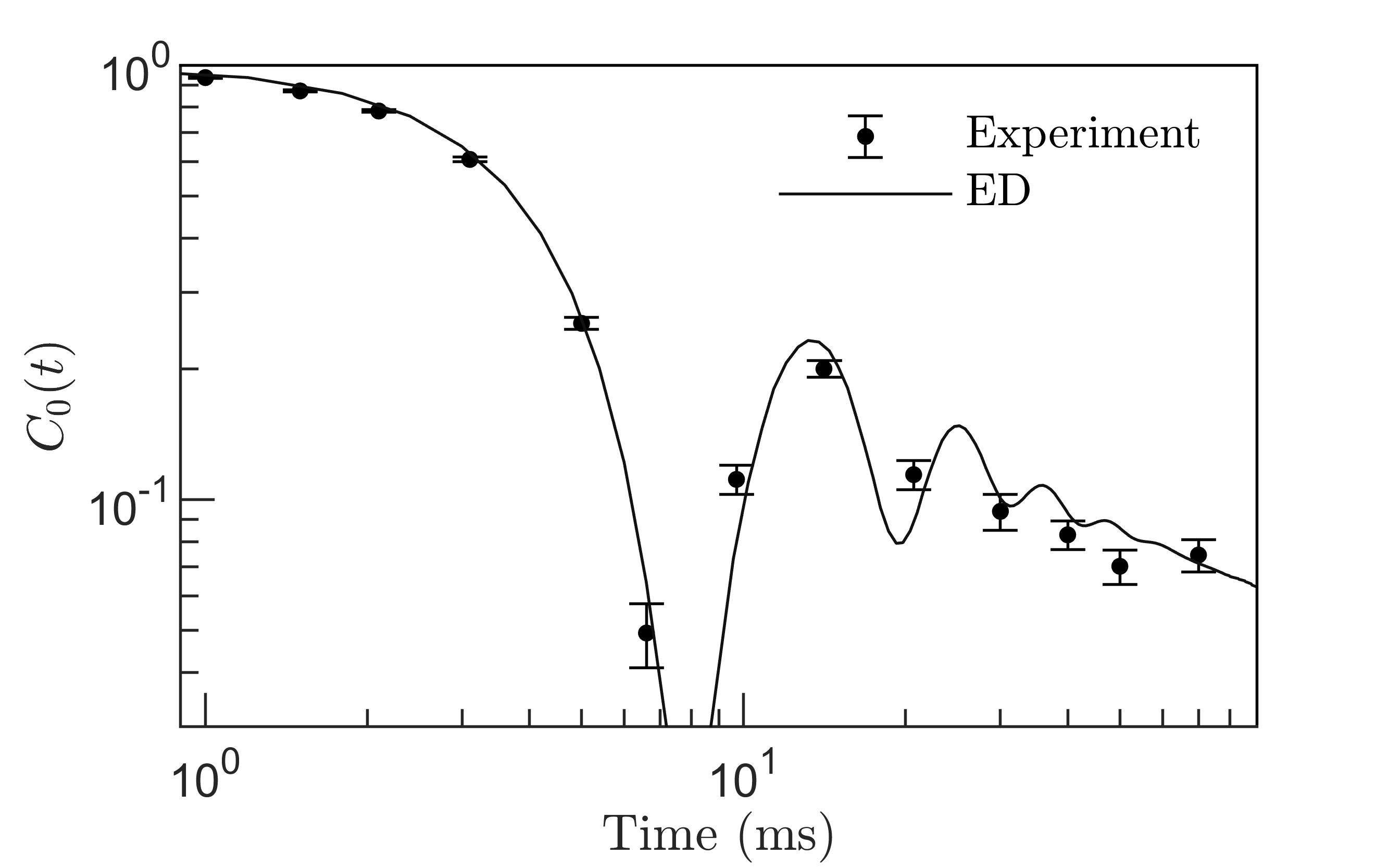}
    \caption{Auto-correlation function from experiment and exact diagonalization (ED). In ED we used the experimental $J_{ij}$ matrix for $\alpha=1.5$ and $L=25$ as well as the $240$ initial states used in the experiment.}
    \label{fig:ED_vs_Exp}
\end{figure}
In Fig.~\ref{fig:ED_vs_Exp} we compare the experimentally measured auto-correlation function with an exact simulation for the 25 ions chain, finding excellent agreement.

\section{Anomalous transport from L\'evy flights}

Transport in the long-range XY model \eqref{eq:powerlawexp}
is described by L\'evy flights~\cite{PhysRevB.101.020416}. 
The Hamiltonian connects states with an up spin at site $i$ and a down spin at site $j$, $\ket{\uparrow_i \downarrow_j}$, with the opposite configuration $\ket{ \downarrow_i \uparrow_j}$. As these states have equal energy, we can use Fermi's golden rule to estimate the transition probability for a spin exchange as
\begin{equation}
    W_{i\rightarrow j} \equiv |\braket{\uparrow_i \downarrow_j|\hat H| \downarrow_i\uparrow_j }|^2 = \frac{\lambda}{|i-j|^{2\alpha}},
\end{equation}
with $\lambda\propto J^2$. Moreover, $W_{i\rightarrow j} =W_{j\rightarrow i} \equiv W_{ij}$. Hence, we can construct a classical Master equation for the spin excitation probability density $f_i\equiv (\braket{\hat \sigma^z_i}+1)/2$,
\begin{equation}
    \partial_t f_i = \sum_j W_{ij} (f_j-f_i),
\end{equation}
where the first and second term on the right hand side are the ``gain'' and ``loss'' terms, respectively. 

We can solve this equation by Fourier transforming to momentum $k$ and taking both a continuum and infinite system size limit while retaining a short-distance cutoff of unity, leading to 
\begin{equation}
    \partial_t f_k = W_k f_k,
    \label{Eq:ME_mom}
\end{equation}
with the transition rates
\begin{equation}
    W_k=\left( -c_\alpha |k|^{2\alpha-1} + \frac{k^2}{3-2\alpha}\right)\lambda,
    \label{Eq:FT_rates}
\end{equation}
where $c_\alpha= -2 \Gamma(1-2\alpha) \sin(\alpha \pi)$ and $\Gamma$ is the Gamma function. Solving Eq.~\eqref{Eq:ME_mom} and Fourier transforming back for an initially localized excitation $f_j=\delta_{j,0}$, we get
\begin{equation}
    f_j(t)=\int \frac{dk}{2\pi} e^{ikj+W_k t}.
    \label{Eq:solME}
\end{equation}
The exact form of the solution crucially depends on the value of $\alpha$. In the following, we discuss the three cases shown in the dynamical phase diagram in Fig. 1, main text.

\textbf{Diffusion for $\alpha \geq 1.5$: }In this case, the second term in Eq.~\eqref{Eq:FT_rates} dominates. Inserting $W_k$ into Eq.~\eqref{Eq:solME}, we find
\begin{equation}
    f_j^{\alpha>1.5}(t)= (D_\alpha t)^{-1/2} G\left( \frac{|j|}{(D_\alpha t)^{1/2}}\right),
\end{equation}
with a diffusive Gaussian profile $G(y)=\exp(-y^2/4)/\sqrt{4\pi}$ and diffusion coefficient $D_\alpha=\lambda/(2\alpha-3)$. The long-range character of the transition rates only shows up in subleading corrections, leading to algebraic tails of the distribution function~\cite{PhysRevE.100.042140, PhysRevB.101.020416}.

\textbf{Superdiffusion for $0.5<\alpha<1.5$: }In this case, we get 
\begin{equation}
    f_j^{0.5<\alpha<1.5}(t)= (D_\alpha t)^{-1/(2\alpha-1)} F_\alpha\left( \frac{|j|}{(D_\alpha t)^{1/(2\alpha-1)}}\right),
\end{equation}
with $D_\alpha=\lambda c_\alpha$. The distribution function $F_\alpha$ is given by the stable symmetric distributions defined by
\begin{equation}
    F_\alpha(y)=\int \frac{dk}{2\pi} e^{iyk-|k|^{2\alpha-1}},
    \label{eq:F_alpha}
\end{equation}
which only has a known elementary solution for $\alpha=3/2$ (Gaussian) and $\alpha=1$ (Lorentzian).

\textbf{Mean-field regime for $\alpha<0.5$: }For $\alpha<0.5$, the system enters the mean-field (or unstable transport) regime. Here, a consistent hydrodynamic description is no longer possible in the thermodynamic limit, as revealed by the diverging transition rates in Eq.~\eqref{Eq:FT_rates} for $k\rightarrow 0$. Instead, the system forms a zero-dimensional ``quantum dot'' rather than a one-dimensional chain, and relaxes to thermal equilibrium instantaneously~\cite{PhysRevB.101.020416,PhysRevLett.110.170603}, due to the non-local nature of the interactions. This behavior can be best understood from the limit of all-to-all interactions $\alpha=0$, allowing to rewrite the Hamiltonian in terms of a single global spin operator $\vec{\hat \sigma}_{\rm global} =\sum_{i =1}^L\vec{\hat \sigma}_i$. In the thermodynamic limit, the diverging spin eigenvalue associated with this operator gives rise to a vanishing relaxation time.

\section{Short time expansion}
Here, we calculate the short time dynamics of 
\begin{equation}
\Tr\left[ \hat \sigma^z_i(t) \hat \sigma^z_j (0)\right]
\end{equation}
for time evolution under the Hamiltonian
\begin{equation}
    \hat H= \sum_{i<j} J_{ij} (\hat \sigma^+_i \hat  \sigma^-_j + h.c.).
\end{equation}
 We expand $\hat \sigma^z_i(t)$ in the Heisenberg picture,
\begin{align*}
&\hat \sigma^z_i(t) = e^{i\hat Ht}\hat \sigma^z_i e^{-i\hat Ht}\\
&= \hat \sigma_i^z +iJt\left[ \frac{\hat H}{J},\hat \sigma_i^z\right]+\frac{(iJt)^2}{2!}\left[ \frac{\hat H}{J},\left[  \frac{\hat H}{J},\hat \sigma_i^z\right]\right]+\mathcal{O}((Jt)^3).
\end{align*} 
When evaluating the commutators and taking the trace, the first order contribution vanishes and we get
\begin{align*}
\Tr[ \hat \sigma^z_i(t)\hat \sigma^z_j (0)] =& t^2 \sum_{k\neq i} J_{ik}^2 \left(\Tr[\hat \sigma_k^z \hat \sigma_j^z]-\Tr[\hat \sigma_i^z \hat \sigma_j^z]\right)\\
+&\Tr[\hat \sigma_i^z \hat \sigma_j^z].
\end{align*}
Furthermore, $\Tr[\hat \sigma_k^z \hat \sigma_m^z]=\delta_{km}$ and hence
\begin{equation}
\Tr[ \hat \sigma^z_i(t)\hat \sigma^z_j (0)] = \bigg\{\begin{array}{lr}
        1-t^2\sum_{i\neq k}J_{ik}^2  & \text{for } i=j\\
        t^2J_{ij}^2 & \text{for } i\neq j\\
        \end{array}.
\end{equation}

In Fig. 3 of the main text we used the explicit $J_{ij}$ matrix determined from the laser parameters in the experiment to obtain the short time expansion.

\begin{figure}
     \centering
     \includegraphics{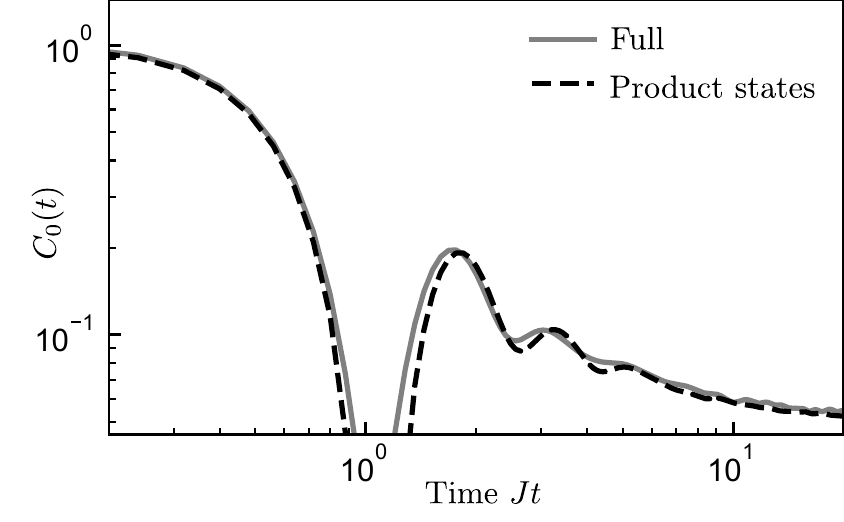}
     \caption{Comparing the evaluation of the full trace with sampling of product initial states. Data obtained from exact diagonalization on an $L=19$ chain with $\alpha=1$. For product state sampling, we used $M=240$ initial states, for full trace we used $10$ Haar-random states.}
     \label{fig:trvsprod}
 \end{figure}
\section{Product state sampling of the trace}

The infinite temperature correlation function in Eq. (3) of the main text can be written as
\begin{align}
    C_j(t)&=\braket{\hat \sigma^z_j(t) \hat \sigma^z_0 (0)}_{T=\infty}\\ &=     \frac{1}{Z} \mathrm{Tr} \left[e^{- \hat H/T}\hat \sigma^z_j(t) \hat \sigma^z_0 (0) \right]\bigg|_{T=\infty}\\ &=     \frac{1}{Z} \mathrm{Tr} \left[\hat \sigma^z_j(t) \hat \sigma^z_0 (0) \right],
\end{align}
with the partition sum $Z=\mathrm{Tr}[e^{- \hat H/T}]|_{T=\infty}=2^L$ for systems of size $L$. We expand the trace in the $\hat \sigma^z$ basis to obtain
\begin{align}
    C_j(t)&= \frac{1}{Z} \sum_{ \sigma_{-\frac{L}{2}}, \dots, \sigma_{\frac{L}{2}}} \braket{\sigma_{-\frac{L}{2}}\dots\sigma_{\frac{L}{2}}|\hat\sigma^z_j(t) \hat \sigma^z_0 (0)|\sigma_{-\frac{L}{2}}\dots\sigma_{\frac{L}{2}}}\\
    &=\frac{1}{Z} \sum_{ \sigma_{-\frac{L}{2}}, \cdots, \sigma_{\frac{L}{2}}} \sigma_0 \braket{\sigma_{-\frac{L}{2}}\cdots\sigma_{\frac{L}{2}}|\hat\sigma^z_j(t) |\sigma_{-\frac{L}{2}}\dots\sigma_{\frac{L}{2}}} \label{eq:infinite_T_allprod} ,
\end{align}
where $\sigma_j = \pm 1$. We have therefore reduced the evaluation of a two-time correlation function to the weighted sum of single-time functions using product initial states, which can be realized in the experiment. In order to evaluate Eq.~\eqref{eq:infinite_T_allprod} exactly, all $2^L$ product states in the z-basis would have to be prepared, time evolved and measured to then obtain $C_j(t)$ from the weighted sum. In practice, however, it suffices to sample a small number $M\approx 10-100$~\cite{PhysRevB.101.020416} of product states to evaluate Eq.~\eqref{eq:infinite_T_allprod} while replacing $Z\rightarrow M$. 
In the main text, we considered the largest sector with magnetization $\sim 0$. In Fig.~\ref{fig:trvsprod} we compare the sampling of the largest sector to the exact results of the thermal expectation value. The full trace is evaluated using the typicality approach~\cite{PhysRevLett.102.110403}, which involves evolving Haar-random states $\ket{\psi_r}$ in the entire Hilbert space and then averaging over $5-10$ such states.  

To further improve the convergence of the sampling of Eq.~\eqref{eq:infinite_T_allprod} we prepare the conjugate configuration for each randomly drawn product state. In the conjugate state, all spins are flipped except for the central one. This choice of initial states reduces statistical noise by reproducing the exact property $C_j(0)=\delta_{j,0}$ of the full trace for each pair of initial states.
\bibliographystyle{apsrev4-1}

\end{document}